\documentstyle[emulateapj,graphicx]{article}
\begin{document}

\title{An energetic blast wave from
the December 27 giant flare of the soft $\gamma$-ray repeater
1806-20}

\author{X. Y. Wang$^{1}$, X. F. Wu$^{1}$, Y. Z. Fan$^{2,3}$,
Z. G. Dai$^{1}$ and B. Zhang$^{2}$}
\affil{ $^1$Department of Astronomy, Nanjing University, Nanjing
210093, China; xywang@nju.edu.cn, xfwu@nju.edu.cn, dzg@nju.edu.cn \\
$^{2}$ Department of Physics, University of Nevada, Las Vegas, NV
89154, USA, yzfan@physics.unlv.edu, bzhang@physics.unlv.edu\\
$^{3}$ Purple Mountain Observatory, Chinese Academy of Sciences,
Nanjing 210008, China}

\begin{abstract}
Recent follow-up observations of the December 27 giant flare of
SGR 1806-20 have detected a multiple-frequency radio afterglow
from 240 MHz to 8.46 GHz, extending in time from a week to about a
month after the flare. The angular size of the source was also
measured for the first time. Here we show that this radio
afterglow gives the first piece of clear evidence that an
energetic blast wave sweeps up its surrounding medium and produces
a synchrotron afterglow, the same mechanism as established for
gamma-ray burst afterglows. The optical afterglow is expected to
be intrinsically as bright as $m_R\simeq13$ at $t\la 0.1$ days
after the flare, but very heavy extinction makes the detection
difficult because of the low galactic latitude of the source.
Rapid infrared follow-up observations to giant flares  are
therefore crucial for the low-latitude SGRs, while for
high-latitude SGRs (e.g. SGR 0526-66), rapid follow-ups should
result in identification of their possible optical afterglows.
Rapid multi-wavelength follow-ups will also provide more detailed
information of the early evolution of a fireball as well as its
composition.
 \end{abstract}


\section{Introduction}
Soft $\gamma$-ray repeaters (SGRs) distinguish themselves from
classical gamma-ray bursts (GRBs) by their repetitive, soft bursts
coming from  nearby sources, likely strongly magnetized neutron
stars dubbed ``magnetars'' (Thompson \& Duncan 1995). Giant flares
 are rare events from SGRs, each characterized by an initial hard spike with more than a
million Eddington luminosity and a pulsating tail persisting for
hundreds of seconds (Mazets et al. 1979; Hurley et al. 1999;
Borkowski et al. 2004). A giant flare was recorded on 2004
December 27
 from SGR 1806-20 (Borkowski et al. 2004; Hurley et al. 2004;
 Mazets et al. 2004; Palmer et al. 2004; Smith et al. 2005).
In comparison with previous ones, this flare is exceptionally
strong with an isotropic energy of gamma-ray emission larger than
$8\times 10^{45}{\rm ergs}$ (Boggs et al. 2005), given a source
distance $d=15.1\rm kpc$ (Corbel \& Eikenberry 2004).

Afterglow emission from giant flares was first detected from SGR
1900+14 at radio frequencies (Frail et al. 1999). The radio
afterglow of the December 27 giant flare from SGR 1806-20 was
detected about 7 days after the flare with a flux density at 8.46
GHz approximately 100 times brighter than the peak flux density of
the SGR 1900+14 radio afterglow (Gaensler et al. 2005; Cameron et
al. 2005). Generally, the multi-frequency radio afterglow light
curves display a power law decay behavior with time during  the
observational time from one week to one month after the flare,
which is quite similar to the afterglows of cosmological GRBs. But
the decay rate at higher frequencies (e.g. 8.46 GHz) is larger
than that at lower frequencies  ({ e.g. 240 MHz}) { and the
spectra clearly show a gradual steepening with time} (Cameron et
al. 2005).

Cheng \& Wang (2003) suggested the possibility that the radio
afterglow of the SGR 1900+14 giant flare might be explained with
the same mechanism as  established for GRB afterglows
(M\'{e}sz\'{a}ros \& Rees 1997), i.e. the emission from the
expanding blast wave that forms when the outflow from the SGR
interacts with the surrounding interstellar medium (ISM). However,
the data of that radio afterglow are too sparse for us to make a
firm conclusion. The radio afterglow of the December 27 flare has
much more abundant data of the light curves which was detected
from 240 MHz to 8.46 GHz in frequency and from a week to a month
after the flare in time. The angular size of the source was also
measured for the first time. In this {\it Letter}, we will show
that this radio afterglow gives the first piece of clear evidence
that an energetic blast wave sweeps up its surrounding medium and
produces a synchrotron afterglow.

\section{The blast wave model for the radio afterglow }
It was  suggested by Huang et al. (1998) that relativistic
fireballs similar to those in classic GRBs may exist in SGR
bursts. Thompson \& Duncan (2001) proposed that the
extraordinarily high peak luminosity $L>10^6 L_{\rm Edd}$ (where
$L_{\rm Edd}$ is the Eddington luminosity), hard spectrum and
rapid variability of the initial spike emission of giant flares
imply that the spike emission must originate from a
relativistically expanding fireball with the initial Lorentz
factor of at least several  in order to avoid the pair production
problem. Such a relativistic outflow, after emitting the hard
spike, should retain some energy and then drives a blast wave
expanding into the surrounding medium.

\subsection{The angular size and constraint on the flare energy}
Now let's first consider the dynamics of a  blast wave that is
driven by the relativistic outflow and  expands into the ISM. The
blast wave  will be gradually decelerated by the swept-up ISM and
enter the non-relativistic phase at $t_{\rm nr}=(3E/4\pi nm_p
c^5)^{1/3}=4.5 E_{46}^{1/3}n_0^{-1/3}\,{\rm d}$, where $E$ is the
isotropic kinetic energy of the blast wave, $c$ is the speed of
light, $n$ is the number density of ISM,
 $m_p$ is the proton mass, and we have used the usual notation
$a\equiv10^{n}a_n$ in c.g.s. units except that the observer's time
$t$ is in unit of day. So, for $E_{46}/n_0<4$, the blast wave must
have entered the non-relativistic phase during the observational
time frame ($t>7\rm d$).
During this non-relativistic phase, the speed (in units of $c$)
and the radius of the blast wave are given by
$\beta=[{12E}/({125\pi n m_p c^5 t^3})]^{1/3}=0.4
E_{46}^{1/5}t_1^{-3/5}n_0^{-1/5}$ and $R={5}\beta c
t/2=2.6\times10^{16}E_{46}^{1/5}t_1^{2/5}n_0^{-1/5} {\rm cm}$
respectively (Cheng \& Wang 2003). The measured angular size at
8.46 GHz of the radio afterglow is $\sim  75$ mas around seven to
ten days after the flare  and has  no noticeable variation in
diameter during this period (Gaensler et al. 2005; Cameron et al.
2005). At a distance of $d=15.1{\rm kpc}$, this angular size
implies a subluminal speed of $\sim 0.35 \rm c$ for the averaged
projected expansion and an apparent radius of
$R_{\bot}=9\times10^{15}{\rm cm}$. So, if the outflow is spherical
at this time (i.e. $R=R_{\bot}$), we obtain the constraint on the
outflow energy, i.e. $E\simeq10^{44}n_0 \,\rm ergs$ {\footnote{
The discrepancy between the isotropic gamma-ray energy and the
kinetic energy of the afterglow may be naturally expected within
the magnetar framework as the prompt hard spike emission taps the
magnetic energy, while the afterglow is only related to the
kinetic energy (Zhang \& Kobayashi 2005). On the other hand, this
discrepancy could be alleviated by a denser environment. }}. If,
instead, the outflow is mildly beamed in the non-relativistic
phase, then the half opening angle $\theta$ satisfies ${\rm sin}
\theta=R_{\bot}/R=0.4 (E_{46}/n_0)^{-1/5}$. In any case, the real
kinetic energy (after beaming correction) of the outflow is about
$10^{44}-10^{45} {\rm ergs}$ for a typical ISM with $n=1{\rm
cm^{-3}}$ and $d=15.1{\rm kpc}$.
 It is more reasonable to believe that
the outflow becomes approximately spherical  at the
non-relativistic phase when the sideways expansion takes place
(Livio \& Waxman 2000). In the following we will focus on the
spherical outflow case with $E=10^{44}n_0 \,\rm ergs$.
The apparent radius should increase with time as $R\propto
t^{2/5}$ in the non-relativistic phase, so it will increase by a
factor of $1.15$ from 7 to 10 days. This is consistent with the
fact that there is  no noticeable observed variation of the
angular size during the period{\footnote{ After the submission of
the paper, more observation data of the angular size at time from
10 to 20 days are reported in Gaensler et al. (2005). The angular
size shows a slow increase with time, being consistent with the
$R\propto t^{2/5}$ relation. }}.

\subsection{The light curves  of the radio afterglow }
As the blast wave sweeps up the surrounding medium, the shock
accelerates electrons and amplifies the magnetic field in the
swept-up matter (Blandford \& Eichler 1987). The afterglow
emission arises from synchrotron radiation of these shocked
electrons in the magnetic field. The electron energy distribution
in the downstream of the shock usually manifests as a single power
law or broken power law. { The radio-band spectra show a gradual
steepening with time, with a single power-law index of
$\beta_1=0.62\pm0.02$ around $t=7 {\rm d}$, while at $t=23 \rm d$,
$\beta_2=-0.91\pm0.08$ (Cameron et al. 2005)}. There are also
reports that a possible break exists in the radio-band spectrum at
early epoches (Gaensler et al. 2005). For example, at about 9 days
after the giant flare, the measured power law spectral index
between 4.8 and 8.6 GHz is $\sim-1.0$, while between 2.4 and 4.8
GHz the index is $\sim  -0.66$ (Gaensler et al. 2005).The
steepening of  the light curve decay with frequency also favors
that there is a spectral break between { 240 MHz} and 8.46 GHz in
the synchrotron blast wave model. This break is unlikely to be the
cooling break because the cooling frequency is much higher than
the radio band (see Eq.2). As suggested by Li \& Chevalier (2001)
to explain the afterglows of GRB991208 and GRB000301C, we consider
that this break is caused by an intrinsic break in the energy
distribution of the shock-accelerated electrons, which takes the
form of
\begin{equation}
\frac{dN_e}{d\gamma_e}= \left\{
\begin{array}{l}
 C\gamma_e^{-p_1}, \,\,\,\,\,\,{\rm if} \,\gamma_{\rm min}<\gamma_e<\gamma_b \\
 C\gamma_b^{p_2-p_1}\gamma_e^{-p_2},     \,\,\,\,\,           {\rm if } \,\gamma_e>\gamma_b
                             \end{array} \right.  \;.
\end{equation}
where $\gamma_{\rm min}$ and $\gamma_b$ are the minimum and break
Lorentz factors respectively. The observed synchrotron spectrum of
the Crab Nebula  supports the intrinsic break in the injected
particle spectrum (Amato et al. 2000). Similarly, we assume that
the ratio of the break energy and the minimum energy,
$R_b\equiv\gamma_b/\gamma_{\rm min}$, remains essentially a
constant in time so that the relative shape of the distribution is
time invariant (Li \& Chevalier 2001). From the values of
$\beta_1$ and $\beta_2$, we infer $ p_1\simeq2.2$ and $p_2\simeq
3.0$.

Assuming that shocked electrons and the magnetic field acquire
constant fractions ($\epsilon_e$ and $\epsilon_B$) of the total
shock energy density, we get $\gamma_{\rm
min}\simeq\epsilon_e\frac{p_1-2}{p_1-1}\frac{m_p}{m_e}(\Gamma-1)$
and $B=({16\pi \epsilon_B \beta^2 n m_p
c^2})^{1/2}=1.4\times10^{-2}
\epsilon_{B,-1}^{1/2}E_{44}^{1/5}t_1^{-3/5}n_0^{3/10}{\rm G}$.
Thus we can obtain the break frequency corresponding to
$\gamma_{\rm min}$, the break frequency corresponding to the
electron break caused by the synchrotron cooling, and the peak
flux of the spectrum, which are given by
\begin{equation}
\begin{array}{ll}
\nu_m=3\times10^5
[\frac{3(p_1-2)}{p_1-1}]^2(\frac{\epsilon_e}{0.3})^2E_{44}
t_1^{-3}n_0^{-1/2}\epsilon_{B,-1}^{1/2} {\rm Hz}\\
\nu_c=1.4\times10^{18}
\epsilon_{B,-1}^{-3/2}E_{44}^{-3/5}n_0^{-9/10}t_1^{-1/5} {\rm Hz}\\
F_{\nu_m}=95E_{44}^{4/5}t_1^{3/5}n_0^{7/10}\epsilon_{B,-1}^{1/2}({d}/{15.1\rm
kpc})^{-2} {\rm Jy}.
 \end{array}
\end{equation}
For a broken power-law electron distribution described by Eq.(1),
the light curves at the frequencies below and above the break
frequency $\nu_b$ (which is the synchrotron frequency
corresponding to $\gamma_b$) are different. According to the
afterglow theory, the light curves in the non-relativistic phase
should have the decay slopes of $(21-15p_1)/10$ and
$(21-15p_2)/10$, respectively. For $ p_1=2.2$ and $p_2=3$, the
predicted slopes are in agreement with the observed ones (i.e. {
$\alpha_1=-1.43\pm0.17$ for  240 MHz and $ \alpha_2=-2.63\pm0.13$
for 8.46 GHz}) of the radio afterglow data ({ Cameron et al.
2005}),  if the break frequency $\nu_b$ locates between 240 MHz
and 8.46 GHz. The 4.86 GHz light curve shows a break around
$t=9{\rm d}$, which implies that, in the above picture, $\nu_b$
just crosses $4.86$ GHz at this time. This then suggests
$R_b=130(\epsilon_e/0.3)^{-1}\epsilon_{B,-1}^{-1/4}E_{44}^{-1/2}n_0^{1/4}$.
The shifting of $\nu_b$ to lower frequencies would make 1.43 GHz
light curve decay faster at late times.

The flux density at 1.43 GHz is
\begin{equation}
 F_\nu={ 0.26}E_{44}^{\frac{5p_1+3}{10}}n_0^{\frac{19-5p_1}{20}}
\epsilon_{B,-1}^{\frac{p_1+1}{4}}
\left(\frac{\epsilon_e}{0.3}\right)^{p_1-1}t_1^{\frac{21-15p_1}{10}}
(\frac{d}{15.1\rm kpc})^{-2}{\rm Jy},
\end{equation}
where the coefficient is calculated for $ p_1=2.2$. At $t=9.87
{\rm d}$, the observed flux density at 1.43 GHz is $96\pm2{\rm
mJy}$ (Cameron et al. 2005). So, for typical parameter values of
$E=10^{44}\rm ergs$, $n=1\rm cm^{-3}$, $d=15.1{\rm kpc}$ and
$\epsilon_e=0.3$, we infer $\epsilon_B=0.04$. This equipartition
factor is close to that inferred for GRB afterglows (e.g. Wijers
\& Galama 1999; {Panaitescu \& Kumar 2001}) and corresponds to a
magnetic field of $\sim 10$ mG in the shocked medium.
Nevertheless, we should note that the observations are all in the
power-law domain and the spectral transitions (e.g. $\nu_m$ and
$\nu_c$) have not been observed, which makes  the physical
parameters of this flare not well constrained.

We have also carried out detailed numerical calculations of the
radio afterglow light curves at five frequencies, i.e. 8.46 GHz,
4.86 GHz, 1.43 GHz, 608 MHz and 240 MHz.
The numerical model contains more careful treatments to the blast
wave dynamics during the trans-relativistic stage (Huang et al.
2000) and the broken power-law electron distribution{\footnote{The
code also takes account of the re-derived  electron distribution
by Huang \& Cheng (2003) for afterglows in the deep Newtonian
phase. }}. It is found that the numerical model can give a
satisfactory fit to the light curve data over five radio
frequencies, as shown in Fig.1. {The theoretic light curve of
610MHz are somewhat above the observation data. This discrepancy
is also evident in the power-law fit of the radio  spectra in
Fig.1 of Cameron et al. (2005).}  The parameter values used in the
fit are consistent with those constrained by the above analytic
study{\footnote{For a smaller source distance $d$ (Cameron et al.
2005) , $E/n$ should be correspondingly smaller in order to be
consistent with the measured angular size. As a result, a larger
value for $\epsilon_B$ is then inferred. However, the global
agreement of the radio afterglow behavior with the blast wave
model remains unchanged. }}. An interesting finding is that,
unlike GRB radio afterglows that usually peak around a week after
the burst, the predicted SGR radio afterglows are significantly
stronger at earlier times. This calls for an even more rapid
follow-up in the radio band.

\section{Optical, infrared and X-ray afterglows of giant flares and intermediate
bursts of SGRs} SGR afterglows have been detected only in the
radio band so far. Within the above picture, it is natural to
expect optical and infrared afterglows as well produced by the
same relativistic fireball accompanying the SGR giant flares.   At
the early time $t<t_{\rm nr}$, the blast wave is relativistic and
its emission spectrum is characterized by
\begin{equation}
\begin{array}{ll}
\nu_m=7.5\times
10^{10}[\frac{3(p-2)}{p-1}]^2(\frac{\epsilon_e}{0.3})^2E_{44}^{1/2}
\epsilon_{B,-1}^{1/2}t_{-1}^{-3/2} {\rm Hz}\\
\nu_c=1.1\times
10^{18}\epsilon_{B,-1}^{-3/2}E_{44}^{-1/2}n_0^{-1}t_{-1}^{-1/2}{\rm
Hz}\\
F_{\nu_m}=15\epsilon_{B,-1}^{1/2}E_{44}n_0^{1/2}({d}/{15.1\rm
kpc})^{-2} {\rm Jy} .
 \end{array}
\end{equation}
The optical flux density at {\it R}-band ($4.55\times 10^{14}{\rm
Hz}$) is
\begin{equation}
F_{R}=30(\frac{\epsilon_e}{0.3})^{p-1}\epsilon_{B,-1}^{\frac{p+1}{4}}
E_{44}^{\frac{p+3}{4}}n_0^{1/2}t_{-1}^{-\frac{3(p-1)}{4}}(\frac{d}{15.1\rm
kpc})^{-2}{\rm mJy} ,
\end{equation}
{where $p\simeq2.4$ is the  power-law index of the electron
distribution inferred from the global spectrum $F_\nu\propto
\nu^{-0.68\pm0.02}$ (Cameron et al. 2005).} For typical parameter
values and the inferred value $ \epsilon_B=0.04$ from the radio
afterglows, the magnitude of the optical afterglow at $t=0.1 \rm
d$ is about $m_{R}\simeq13 $. Considering that, for an initial
beamed outflow, the isotropic equivalent energy in the
relativistic phase may be larger than that in the non-relativistic
spherical phase, the above estimate is conservative. Because the
characteristic synchrotron frequency $\nu_m$ at early times is
generally below the optical frequency for an initial Lorentz
factor of $\Gamma_0\sim10$, the optical afterglow reaches its peak
at the deceleration time of the relativistic outflow, which is
$t_{\rm dec}=90E_{44}^{1/3}n_0^{-1/3}\Gamma_{0,1}^{-8/3}{\rm s}$.
At this time,
$\nu_m=1.1\times10^{14}({\epsilon_e}/{0.3})^2\epsilon_{B,-1}^{1/2}
n_0^{1/2}\Gamma_{0,1}^4 \,{\rm Hz}$, and the optical afterglow has
a flux density of
\begin{equation}
F_{R}=5(\frac{\epsilon_e}{0.3})^{p-1}E_{44}\epsilon_{B,-1}^
{\frac{p+1}{4}}n_0^{\frac{p+1}{4}}
\Gamma_{0,1}^{2(p-1)}(\frac{d}{15.1\rm kpc})^{-2} {\rm Jy}.
\end{equation}
Such a bright optical afterglow would be easily detected by rapid
response detectors, such as Ultraviolet/Optical Telescope (UVOT)
on {\it Swift} and Robotic Optical Transient Search Experiment
(ROTSE), if the optical extinction to the source is not very
large. However, SGR 1806-20 has a low Galactic latitude and has
optical extinction of $A_V\sim 30 \,{\rm mag}$ (Eikenberry et al.
2001). This makes it impossible to observe the optical afterglow
from this SGR. The difficulty also applies for the other two SGRs
at low Galactic latitudes, SGR 1900+14 and SGR 1627-41, whose
extinction is also large (Vrba et al. 2000; Wachter et al. 2004).
The {\it R}-band extinction of SGR 1900+14 is moderate , i.e.
$A_R\sim 7\, \rm mag$ (Akerlof et al. 2000), leaving some hope for
optical afterglow detection at its peak. The best target for
optical follow-up observations among the four known SGRs is SGR
0526-66, which lies in the Large Magellanic Cloud and has
$A_V\simeq 1$. At a distance of $d\simeq 50 \,\rm kpc$, the
optical afterglow of a similar giant flare from this SGR is
expected to be bright enough for follow-ups. Even the less rare
intermediate bursts
 with the isotropic energies ranging from
$10^{41}$ to $10^{43}$ ergs (e.g. Golenetskii et al. 1984;
Guidorzi et al. 2004) are expected to have optical afterglows
brighter than $m_R\simeq 23$ at $t<0.1 {\rm d}$ for SGR 0526-66.
This is well above the {\it Swift} UVOT sensitivity. Detecting
such an optical afterglow would further test the blast wave model
and provide more detailed information of the early evolution of a
fireball.

Because of the high optical extinction of low-latitude SGRs,
infrared observations to giant flares and intermediate bursts
appear very useful. The {\it K} band flux density would be,
according to Eq. (6), $F_{\rm 2.2\mu m}\sim {3} {\rm mJy} $ or
$m_{\it K}\simeq{13} $ at $t= 1{\rm d}$ for typical parameter
values to model the radio afterglow of the December 27 giant
flare. Such a bright infrared afterglow could be well detected by
ground-based infrared telescopes. An example of the (unabsorbed)
light curves of the infrared and optical afterglows calculated
with the numerical model for the radio afterglows is also shown in
Fig.2. The flux of the X-ray afterglow  can be estimated in the
same way. For the above typical parameter values, we find the flux
at $t=0.1{\rm d}$ is about $F_{\rm X}\simeq2.6\times10^{-10} {\rm
erg cm^{-2} s^{-1}}$. This flux is, however, much dimmer than the
observed  X-ray flux of the SGR 1900+14 giant flare (Woods et al.
2001), which is thought to be originated from the neutron star
surface immediately after the burst.

\section{Discussions }
We have shown that the radio afterglow behavior of the December 27
giant flare of SGR 1806-20, including the angular size and
multi-frequency light curves, provides a first  piece of clear
evidence that the fireball/blast wave model indeed works in SGRs
as well. {The polarization emission detected in the radio
afterglow (Gaensler et al. 2005; Cameron et al. 2005)
 at a comparable
level ($1\%-3\%$)  to GRB afterglows  (Covino et al. 1999; Wijers
et al. 1999) is also consistent with the blast wave model.} The
initial Lorentz factor of the outflow that drives the blast wave
is not well constrained, because only the late-time radio
afterglow is available up to now. The composite of the
relativistic outflow is also unknown and could be normal baryon
matter as well as relativistic electron-positron dominated wind
since both could drive relativistic blast waves when they are
interacting with  the surrounding medium.

 {\acknowledgments
XYW would like to thank Y. F. Huang, D. M. Wei, Z. Li and
 T. Lu for useful discussions.
We also thank S. Kulkarni and B. Gaensler for informative
communications. This work was supported by the Special Funds for
Major State Basic Research Projects, the National 973 Project
(NKBRSF G19990754), the National Natural Science Foundation of
China under grants 10403002, 10233010 and 10221001. B. Zhang  is
supported by NASA NNG04GD51G and a NASA Swift GI (Cycle 1)
program.}

\begin{figure*}
\plotone{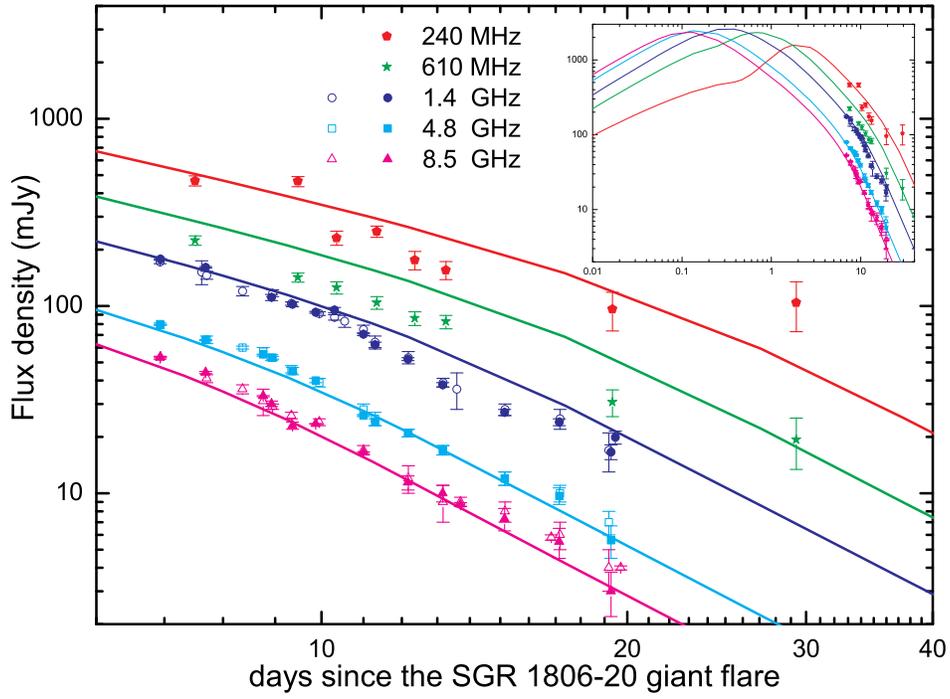} \vspace{0cm} \caption{ Numerical fits of the
radio afterglows of the December 27 giant flare from SGR 1806-20
with the blast wave model. The data are taken from  Gaensler et
al. (2005) (open symbols) and Cameron et al. (2005) (solid
symbols). The parameter values used in the fits are $E=10^{44}{\rm
erg}$, $n=1{\rm cm^{-3}}$, $\Gamma_0=10$,  $\epsilon_{e}=0.32$,
$\epsilon_{B}=0.1$, $p_{1}=2.2$, $p_{2}=3.2$, {$R_{b}=150$} and
$d=15.1 \rm kpc$. The inset shows these light curves over a longer
time range. }
\end{figure*}

\begin{figure*}
\plotone{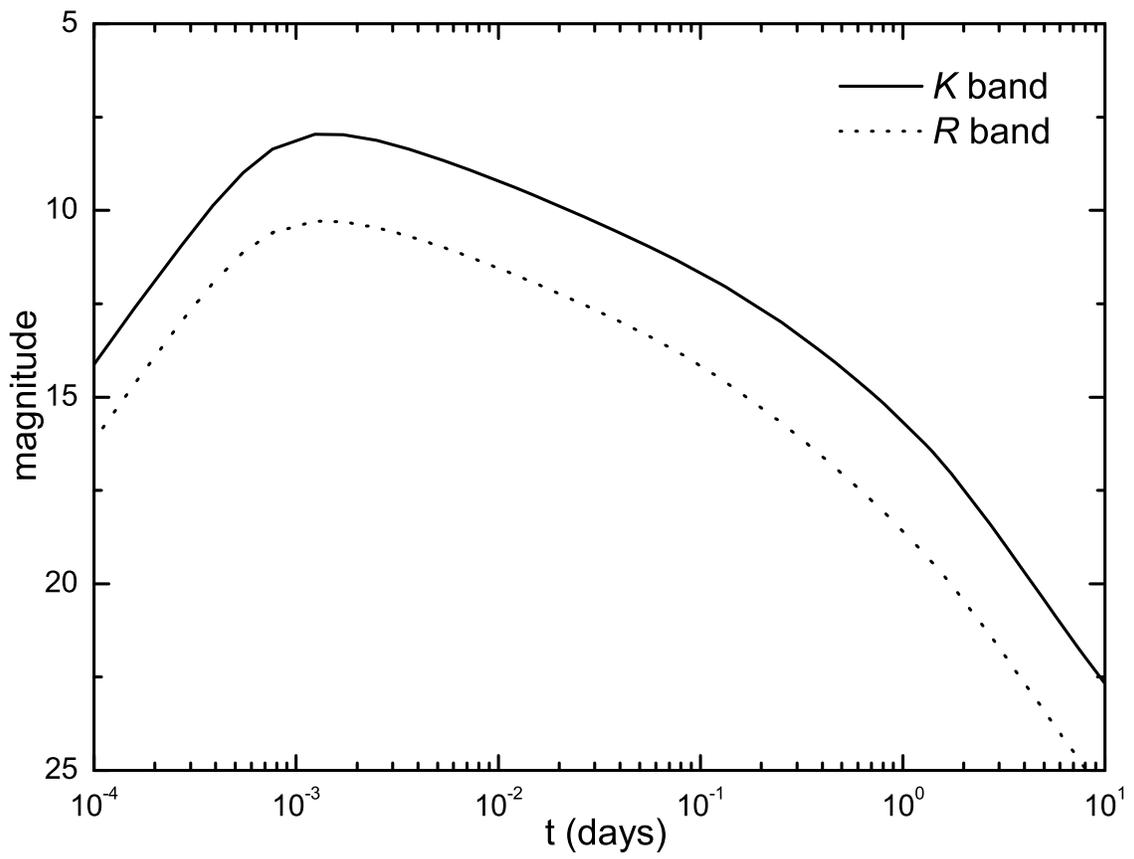} \vspace{0cm} \caption{The predicted optical ({\it
R}-band) and infrared ({\it K}-band) afterglows of a giant flare
similar to the December 27 event. Parameter values used are the
same as those in the fits of Fig. 1. }
\end{figure*}

\end{document}